\documentclass[aps, pre, floatfix,
twocolumn
]{revtex4-1}

\usepackage{graphicx}
\usepackage{bm}
\usepackage{subfigure}
\usepackage{hyperref}
\usepackage{amsmath}
\usepackage{color}

\newcommand{\blue}[1]{{\color{black} #1}}

\begin{document}

\title{Random sequential adsorption of particles with tetrahedral symmetry}

\author{Piotr Kubala}
  \email{pkua.log@gmail.com}
\author{Micha\l{} Cie\'sla}
  \email{michal.ciesla@uj.edu.pl}
\affiliation{M.\ Smoluchowski Institute of Physics, Department of Statistical Physics, Jagiellonian University, \L{}ojasiewicza 11, 30-348 Krak\'ow, Poland}
\author{Robert M. Ziff}
    \email{rziff@umich.edu}
\affiliation{Center for the Study of Complex Systems and Department of Chemical Engineering, University of Michigan, Ann Arbor MI 48109-2136, U. S. A.}

\date{\today}

\begin{abstract}
We study random sequential adsorption (RSA) of a class of solids that can be obtained from a cube by specific cutting of its vertices, in order to find out how the transition from tetrahedral to octahedral symmetry affects the densities of the resulting jammed packings. We find that in general solids of octahedral symmetry form less dense packing, however, the lowest density was obtained for the packing build of tetrahedra. The densest packing is formed by a solid close to a tetrahedron but with vertices and edges slightly cut. Its density is $\theta_{max} = 0.41278 \pm 0.00059$ and is higher than the mean packing fraction of spheres or cuboids but is lower than that for the densest RSA packings built of ellipsoids or spherocylinders. The density autocorrelation function of the studied packings is typical for random media and vanishes very quickly with distance.

\end{abstract}

%
\maketitle
\section{Introduction}
Random packings and their properties is a very active field of research as they can successfully model a wide range of structures appearing in nature, including granular, soft and bio-matter. Random packings are studied to explain some processes important from a utilitarian point of view, like self-assembly \cite{Whitesides2002}, glass formation \cite{Truskett2000} or adsorption \cite{Dabrowski2001}. 

Properties of random packings may be sensitive to the particular shape of the solids that built them, however, what seems to be the key factor is the symmetry of the shape \cite{Wouterse2007, Jiao2008, Jiao2010}. For example, for so-called close packings, where neighboring objects are in contact, it is expected that solids of higher symmetry form less dense packings \cite{Baule2013, Baule2014}. This has been confirmed for several three-dimensional solids like, for example, ellipsoids \cite{Donev2004, Man2005} or spherocylinders \cite{Zhao2012, Ferreiro2014, Meng2016}, which can form denser packing than spheres.

In this study, we want to check whether the same is valid for solids of symmetry described by point groups, i.\,e., if the packing fraction is lower for objects with a larger, but finite number of symmetries. The focus is on the transition between tetrahedral to octahedral symmetry. Jammed packing composed of such solids have been studied recently in the context of dense packing and self-assembly \cite{Chen2014, Du2017, Klotsa2018, Teich2019}, and it appears that, in general, regular structures appear at lower densities when octahedral symmetry is present \cite{Klotsa2018}. 

As a model of random packing we use random sequential adsorption (RSA) \cite{Feder1980, Evans1993}, which appears to be the simplest yet non-trivial model of random packing that accounts for excluded volume effects. In contrast to the more commonly studied random close packings \cite{Shui2008, Torquato2010, Jaoshvili2010, Agarwal2011}, the packings formed by RSA have well-defined mean densities \cite{Torquato2000, Asencio2017}, and are also more straightforward to carry out.

RSA packings are formed by adding successive objects of a given shape at random position and orientation, but the newly added object cannot intersect with any of previously added objects. If there is an intersection, the new object is removed from the system and abandoned. When placed, the object is not allowed to change its position or orientation. This arrangement of objects has most physical applications in two-dimensional space, where RSA packings resemble monolayers obtained in irreversible adsorption processes \cite{Feder1980}. Properties of such packings are also studied in higher dimensions \cite{Jodrey1980, Sherwood1997, Bonnier2001}, where there may be no direct physical process that can form them, but they are of theoretical interest for various reasons \cite{Widom66,BlaisdellSolomon82,Penrose01}.  It is interesting to compare RSA with random close packings, which of course can be created experimentally.  RSA configurations can also be used as initial states of Monte-Carlo or molecular dynamics particle simulations, and for these various reasons it is useful to investigate their properties.
\section{Model}
\begin{figure}[htb]
  \centering
  \includegraphics[width=0.9\columnwidth]{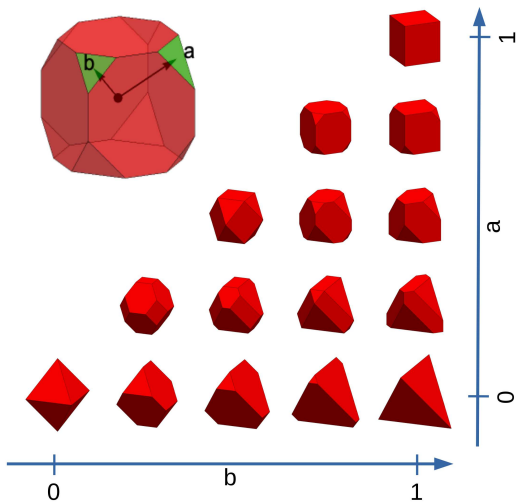}

  \caption{Examples of studied solids. Vectors $\vec{a}$ and $\vec{b}$ originating in the center of the cube point to cutting planes. The length of these vectors corresponds the amount left after cutting. Faces cut by planes pointed by vectors $\vec{a}$ and $\vec{b}$ are green while all other faces are red. Parameters $a$ and $b$ denote the amount left after cutting and are normalized to be in the $[0,1]$.}
  \label{fig:model}
\end{figure}

All solid objects studied here can be obtained from a cube by specific cutting of its vertices. Each vertex is cut by a plane which is perpendicular to the line passing through the center of the cube and this vertex. The maximal possible cut is the one for which the plane contains all three neighboring vertices of the cut vertex. Additionally, all the vertices are divided into two distinct sets in a way that vertices connected by a cube's edge are in different sets, and all vertices in the same set are cut equally---see Fig.\ \ref{fig:model}.
The shape of the solid is fully determined by the length of two vectors $\vec{a}$ and $\vec{b}$ pointing at the cut planes. Assuming that the cube edges have unit length, the lengths of these vectors are in the interval $[\sqrt{3}/6, \sqrt{3}/2]$. Therefore if $a \equiv \sqrt{3}|\vec{a}| - 1/2$ and $b \equiv \sqrt{3}|\vec{b}| - 1/2$, the shape of the solid is described by parameters $a$ and $b$, where both are taken from $[0, 1]$, and $(a, b)$ and $(b, a)$ describe the same object. Coordinates of vertices and volumes are given in Appendix A.  Here we are using the notation similar to that of Ref. \cite{Klotsa2018, Teich2019} to describe these shapes.

Example solids are presented in Fig.\ \ref{fig:model}. Parameters $a=1$ and $b=1$ correspond to a cube, while $a=0$, $b=1$ or $a=1$, $b=0$ describe a tetrahedron. When both these parameters are zero, we get an octahedron, while $a=b=1/2$ describes the cuboctahedron. By construction, all these solids possess tetrahedral symmetry.

We studied random packings for over $100$ different objects described by parameters $(a, b)$. To determine packing properties, $100$ independent random packings were generated according to the RSA algorithm. The packing generation was stopped when the number of iterations exceeded $t_{max} V/V_s$, where $V_s$ is a volume of a single solid, $V$ is a volume of the whole packing, and $t_{max} = 10^8$. In our simulations, solids were scaled to have $V_s=1$ and they were placed into a cube of a volume of $V=2.7 \times 10^4$. We define the dimensionless time $t$ by
\begin{equation}
    t = \frac{n V_s}{V},
\end{equation}
where $n$ is the number of trials.
The dimensionless time is commonly used to compare the number of iterations needed to create different size packings, as it scales with both the volume of the packing as well as the volume of the packed particles. The total number of iterations should be as high as possible in order to create packings which are close to the saturated state, i.e., when there is no possibility to place another, non-intersecting solid in it. Based upon of previous studies of RSA of three-dimensional objects \cite{Sherwood1997, Ciesla2018cubes, Kubala2018, Ciesla2018cuboids, Ciesla2019}, we chose $t_{max}=10^8$ as a good compromise between simulation accuracy and computational time.  To reduce finite-size effects, periodic boundary conditions were used \cite{Ciesla2018}.
The intersection between solids was tested using the separating axis theorem, which was discussed in more detail in \cite{Kubala2018}. 
During packing generation we measured the dependence of the mean packing fraction  on the number of iterations,
\begin{equation}
    \theta(t) = \langle N(t) \rangle \frac{V_s}{V},
\end{equation}
where $\langle N(t) \rangle$ is the mean number of solids in a packing after the number of RSA iterations corresponding to $t$. The averaging is over different, independently generated packings.
\section{Results and discussion}
\begin{figure}[htb] 
  \centering
  \subfigure[]{\includegraphics[width=0.4\columnwidth]{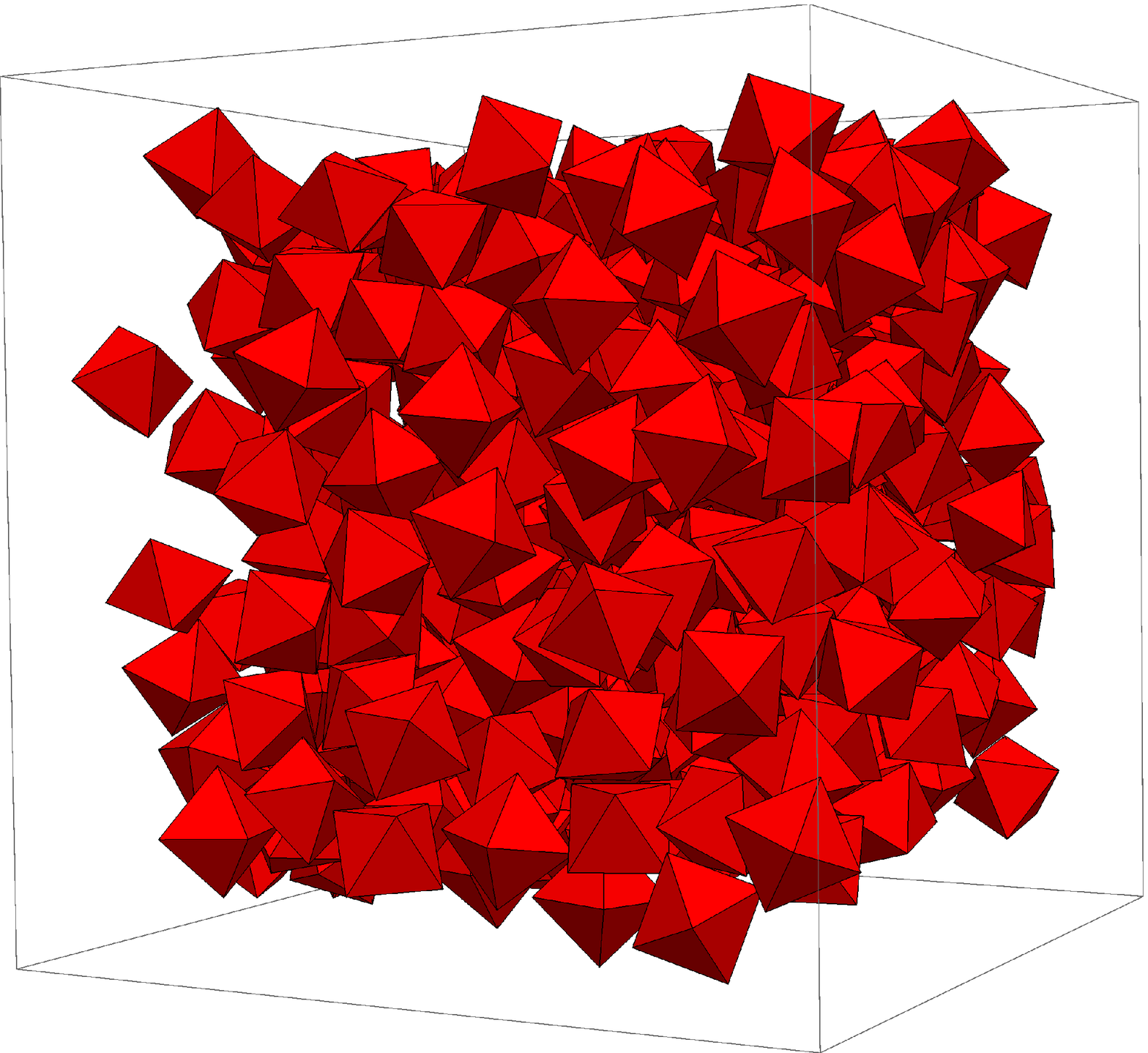}} \subfigure[]{\includegraphics[width=0.4\columnwidth]{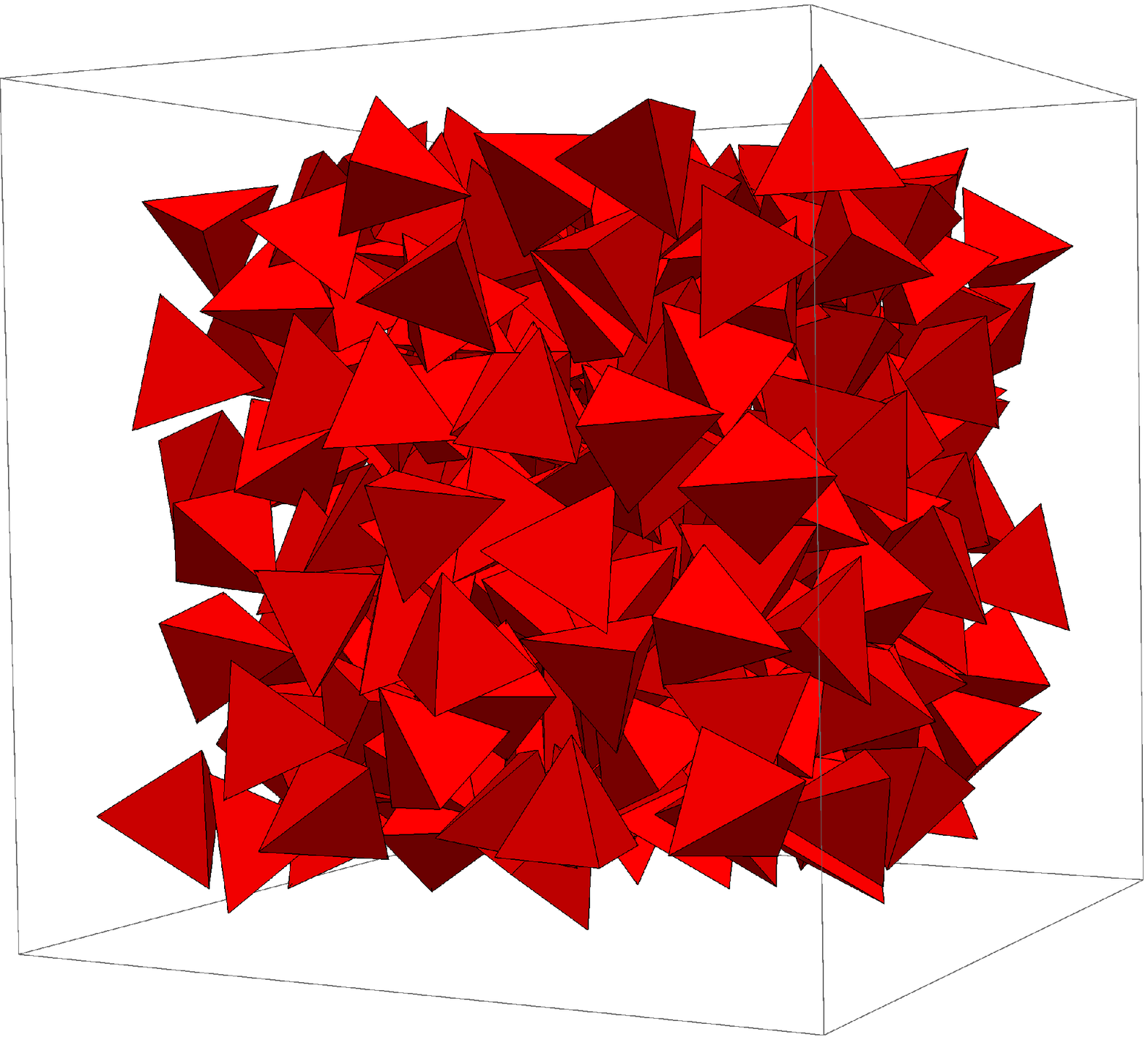}}
  \subfigure[]{\includegraphics[width=0.4\columnwidth]{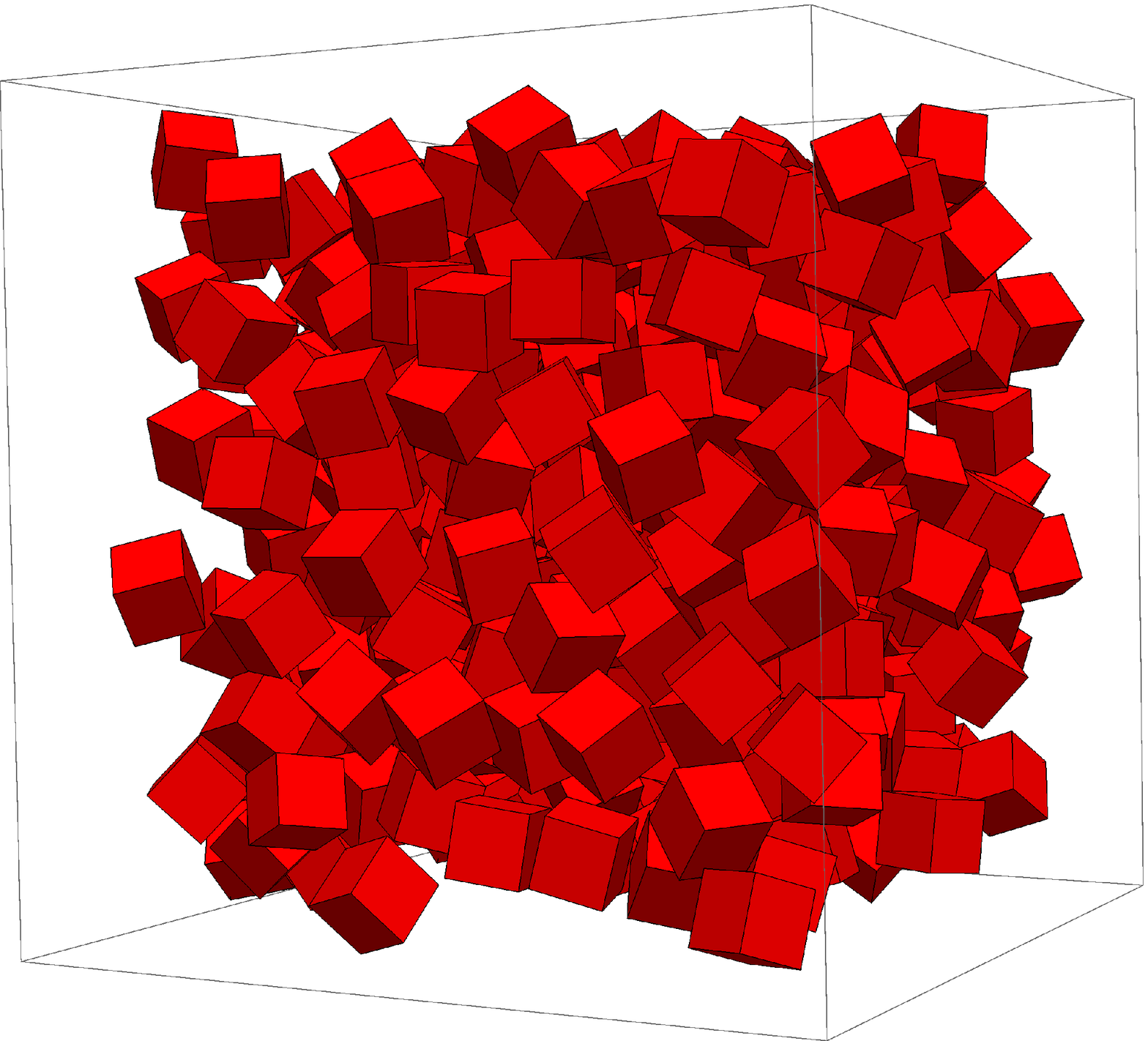}} \subfigure[]{\includegraphics[width=0.4\columnwidth]{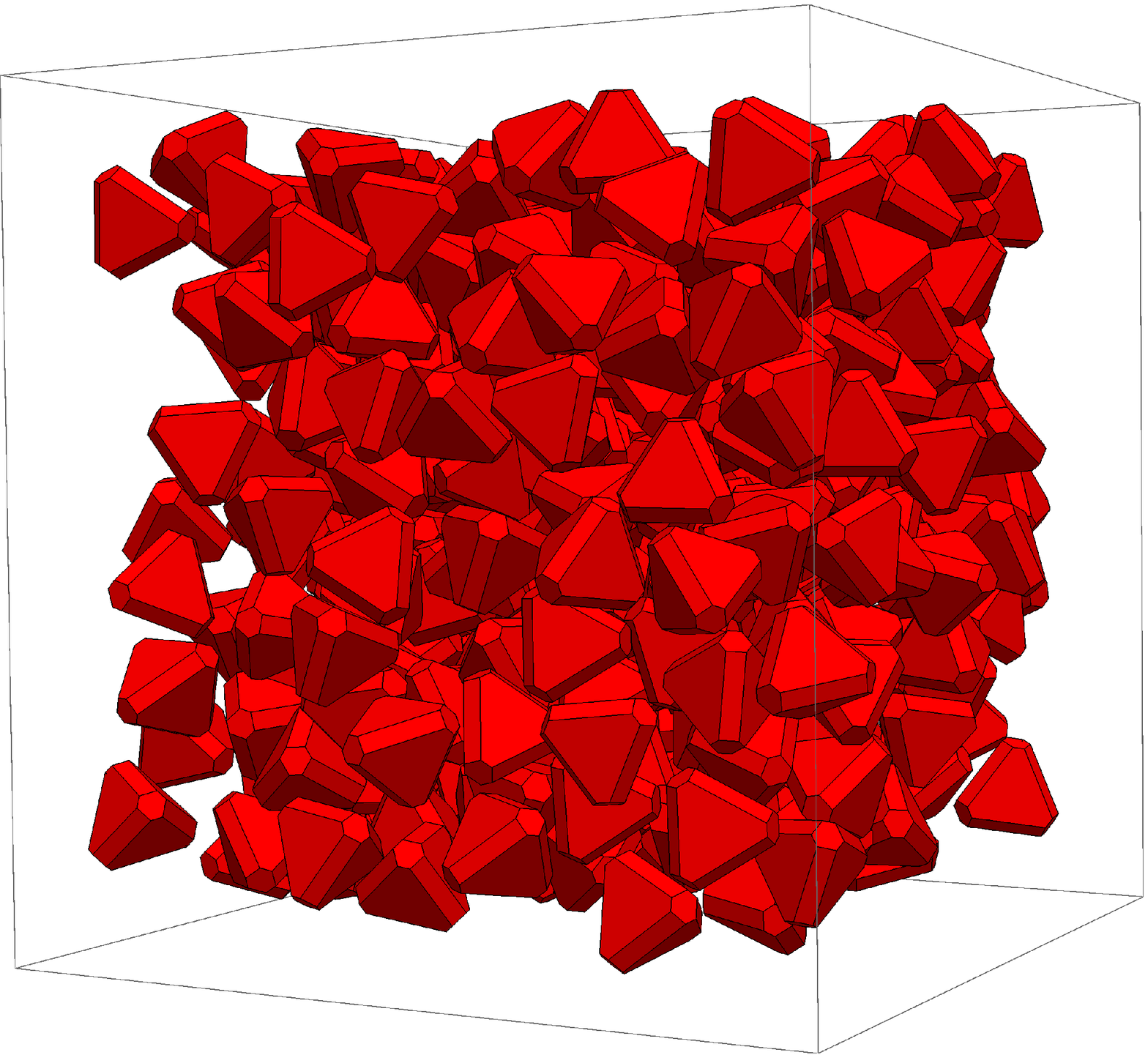}}
  \caption{Fragment of example packings of solid defined by a several $a$ and $b$ values. Panels (a), (b), and (c) show RSA packings build of octahedrons ($a=0,\, b=0$), tetrahedrons ($a=0,\, b=1$) and cubes ($a=1,\, b=1$), respectively. Panel (d) shows packing build of solids that forms densest RSA packings ($a=0.15,\, b=0.65$).}
  \label{fig:example}
\end{figure}

Example RSA packings are shown in Fig.\ \ref{fig:example}.
 The number of RSA iterations needed to get saturation has been studied for packing of spheres, where there exists a method that allows one to generate completely saturated packings \cite{Wang1994, Zhang2013}, and for similarly sized packings it behaves like a random variable of a power-law distribution with median around $10^{11} V/V_s$ \cite{Ciesla2017}, which is a few orders of magnitude higher than the limit used here. Therefore, to get the packing fraction of the saturated state, an extrapolation using packing growth kinetics is needed. For spherically symmetric particles of sufficiently large packing size and number of iterations, it has been proven that the packing fraction $\theta(t)$ approaches saturation density $\theta$ according to the following power-law:
\begin{equation}
   \theta(t) =  \theta - A t^{-{1}/{d}},
    \label{fl}
\end{equation}
where $t$ is the number of iterations or dimensionless time, and $A$ is some positive constant \cite{Pomeau1980, Swendsen1981}. The parameter $d$ is equal to the packing dimension, but for anisotropic objects it is commonly interpreted as the number of degrees of freedom of the object \cite{Hinrichsen1986}.

\begin{figure}[htb]
  \centering
  \includegraphics[width=0.7\columnwidth]{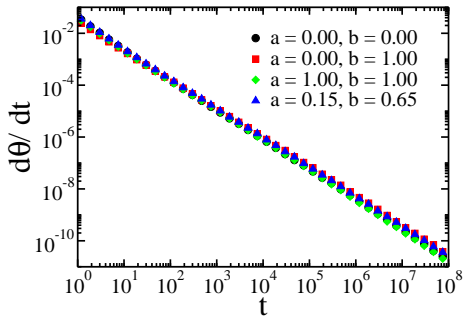}
  \caption{Dependence of the increase of the number of solids in a packing on dimensionless time for several different objects.}
    \label{fig:kinetics}
\end{figure}

The first step to estimate saturated packing fraction $\theta$ from the dependence $\theta(t)$ is determining the exponent $1/d$. Because $\ln [ d \, \theta(t)/dt ] = \ln(A/d) - (1/d+1) \ln t$, the exponent $1/d$ can be obtained from a linear fit to the points $\left(\ln [d\theta(t)/dt],\, \ln t\right)$. The results presented in Fig.\ \ref{fig:kinetics} show that the relation (\ref{fl}) appears to be fulfilled for the studied solids, with the parameter $d$ varying between $5$ and $6.5$. Here, the value of $d$ was estimated from data for $t \in [10^6, 10^8]$.

After having determined $d$, using the substitution $y = t^{-1/d}$, the relation (\ref{fl}) takes the form $\theta(y) = \theta - A y$, which allows one to get the value of $\theta = \theta(\infty)$ from another linear fit.

Applying this procedure to all the studied solids, we determined the saturated packing fractions. Their values are approximately 2\%--3\% higher than the packing fraction at the end of the packing simulations. All numerical values of the presented data are listed in attached files in Supplementary Materials and also in Table \ref{tab:results} in Appendix B.  

\begin{figure}[htb]
  \centering
  \includegraphics[width=0.7\columnwidth]{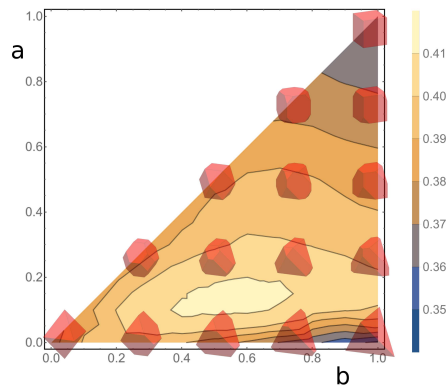}
  \caption{The dependence of mean packing fraction $\theta$ on solid shape. The horizontal axis gives values of $b$ and the vertical axis gives values of $a$.  The values of $\theta$ are given in the legend on the right-hand side.}
    \label{fig:packing-fraction}
\end{figure}

\begin{figure}[htb]
  \centering
  \includegraphics[width=0.7\columnwidth]{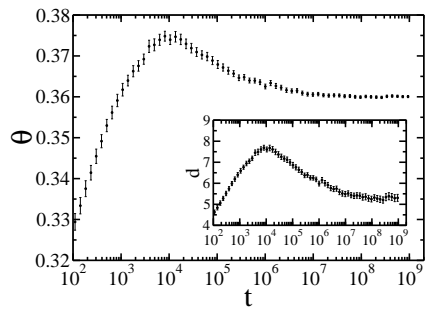}
  \caption{The dependence of the estimation of mean packing fraction $\theta$ for cubes on dimensionless time at which packing generation is stopped. Inset shows the dependence of the estimated value of parameter $d$ on the same variable.}
    \label{fig:kinetics_cubes}
\end{figure}

The saturated packing fraction as a function of parameters $a$ and $b$ is shown as a contour plot in Fig.\  \ref{fig:packing-fraction}. Remembering that octahedral symmetry is present for solids with $a = b$, it can be seen that, indeed, these solids form looser packing than their neighbors in $(a,b)$ space of broken octahedral symmetry $(a \ne b)$. On the other hand, the loosest packing is observed for tetrahedra: $\theta_4 = 0.34750 \pm 0.00049$. For cubes, the RSA packing fraction is slightly larger $\theta_6 = 0.36030 \pm 0.00027$, which improves upon our previous values of $0.3686\pm 0.0015$ \cite{Ciesla2018cubes} and $0.36306\pm 0.0006$ \cite{Ciesla2018cuboids}. All these results are similar, though they differ significantly more than their combined error bars.  We determined that this is due to a systematic error affecting the estimation of the parameter $d$ from the power-law (\ref{fl}) in Refs.\ \cite{Ciesla2018cubes,Ciesla2018cuboids}. Although the slopes observed in Fig.\ \ref{fig:kinetics} seem to be constant, they in fact change slightly with increasing time, and this affects the estimation of the parameter $d$ and consequently the estimate of the mean packing fraction $\theta$. This is shown in Fig.\ \ref{fig:kinetics_cubes}, in which we used data obtained from packings built of cubes generated up to $t=10^9$.
The differences between the obtained packing fractions for cubes become clear when we observe that in previous reports packing generation was stopped at $t=10^5$ \cite{Ciesla2018cubes} and $t=10^6$ \cite{Ciesla2018cuboids}. According to the plot \ref{fig:kinetics_cubes}, the estimation of mean packing fraction for cubes does not changes significantly for $t>10^8$. Thus, the result provided here should be free from the above-mentioned systematic error, but to be certain of that, one would have to develop an algorithm allowing one to generate strictly saturated packings, which has not been done yet.    

\begin{figure}[htb]
    \vspace{0.5in}
    \centering
	\includegraphics[width=0.7\columnwidth]{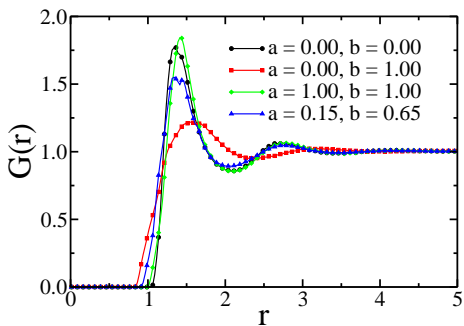}
    \caption{Density autocorrelation functions for four of the studied solids.}
    \label{fig:correlations}
\end{figure}

The densest packing among the platonic solids studied here is built of octahedra: $\theta_8 = 0.37643 \pm 0.00033$. The given errors originate mainly from the uncertainty of the kinetics fit (\ref{fl}), which was approximated using an exact differential. The other, approximately four times smaller source of error, is the statistical error. The finite-size effects are also negligible. As shown in \cite{Ciesla2018}, in the case of packings with periodic boundary conditions, these errors are proportional to oscillations of the density autocorrelation function at a distance comparable to packing size. As we show further (see Fig. \ref{fig:correlations}), these oscillations vanish exponentially fast and are practically not observed for distances significantly smaller than the linear size of the packing used in this study. Therefore, the obtained packing fractions are not affected by this kind of systematic error.

The densest packing overall is observed for a solid described by $a = 0.15$ and $b = 0.65$ with its packing fraction equal to $\theta_{max} = 0.41278 \pm 0.00059$. The solid that forms it is quite similar to a tetrahedron, but its corners are slightly cut and edges slightly shaved; see Fig.\ \ref{fig:example}d. This density is a little smaller than that obtained in the densest RSA packings of ellipsoids or spherocylinders \cite{Sherwood1997, Ciesla2019}, but is significantly higher than the one observed for cuboids \cite{Kubala2018, Ciesla2018cuboids} or spheres \cite{Zhang2013}.

\begin{figure}[htb]
  \centering
  \includegraphics[width=0.7\columnwidth]{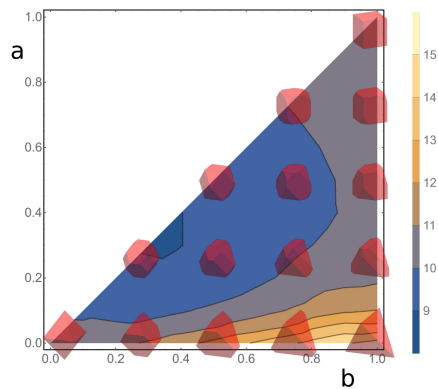}
  \caption{The coefficient $c_1$ from relation (\ref{avf}) for all studied solids, determined from RSA simulations.  The values agree with the theoretical predictions of Eqs.\ (6-13) within a few percent.  The horizontal axis gives values of $b$ and the vertical axis gives values of $a$.}
    \label{fig:c1}
\end{figure}

The difference between the obtained packing fractions can be at least partially explained using the concept of the available volume function, which describes the dependence of the ratio of a volume where subsequent solids can be placed to the actual packing density. For empty packings this function is equal to $1$ and it gets lower with a growing number of objects in a packing until it reaches $0$ for a saturated packing. For small packing densities the available volume function can be expanded into series:
\begin{equation}
    F(\theta) = 1 - c_1 \theta + c_2 \theta^2 + o(\theta^2).
    \label{avf}
\end{equation}
The expansion coefficients depend upon the shape of the solid object and are closely related to virial coefficients $b_n$; for example $c_1$ = $2 b_2$ \cite{Ricci1992}. The available volume function can be estimated during RSA packing generation because it is equal to the probability of successfully adding randomly located and oriented solids in the packing.  The parameter $c_1$ describes the mean volume blocked by a single solid of a unit volume from the perspective of the center of another object. It is worth noting that for the studied platonic solids, $c_1$ is highest for tetrahedra and the lowest for octahedra (see Appendix A). However, the saturation density is determined by all the coefficients, so in general, the relation between the saturated packing fraction $a$ on $c_1$ is not straightforward.  This is illustrated in Fig.\ \ref{fig:c1}, where the value of $c_1$ is shown for all studied solids.

Comparing the results with recent results of Teich et al.\ \cite{Teich2019}, one can notice that the solids leading to the densest random packings fractions are also the ones for which crystallization appears at relatively high densities or it is not observed at all. Teich et al.\ concluded that this is due to competition between different crystal structures. Results presented here suggest that the fundamental reason for this behavior is just exclusion volume effects. This is in line with the observation that the highest densities of random close packings and RSA packings are observed for similar objects \cite{Donev2004, Man2005, Ciesla2019}.

Microstructural properties of the obtained packings can be studied using the two-point correlation function defined as follows:
\begin{equation}
    G(r) = \lim_{dr \to 0} \frac{N(r, r+dr)}{4\pi r^2 \theta \, dr},
\end{equation}
where $N(r, r+dr)$ is the mean number of particles with the center at a distance between $r$ and $r+dr$ from the center of a reference particle. The packing fraction $\theta$ in the denominator is for normalization: $G(r\to \infty) = 1 $. Example results are shown in Fig.\ \ref{fig:correlations}.
As expected, the obtained autocorrelations are typical as for disordered media. The first and highest maximum corresponds to the mean distance of the nearest neighbors. Oscillations of $G(r)$ vanish very fast, which is typical for RSA packings \cite{Bonnier1994}. Indeed, as noted earlier, this fast decay makes the finite-size effects in packings of a linear size of $30$ negligible.

\blue{As the objects analyzed in this study are anisotropic, one can also be interested in how the propagation of orientational order is affected by the change of parameters $a$ and $b$.  In order to measure this quantitatively, polyhedral order parameters can be used \cite{Kubala2019}. They are normalized in such a way that 0 value means complete disorder in orientations, while 1 value means full order. One needs a single order parameter for all regarded polyhedra to make the comparison possible. As all solids studied possess tetrahedral symmetry, the tetrahedral order parameters $\rho_4$ would seem to be the best choice, however it does not capture all of the symmetries of octahedral particles, so a fully oriented sets of those would not yield the maximal value of $\rho_4$. The possible solution is to use octahedral order parameter $\rho_8$, which would recognize the full order for both octahedral and tetrahedral particles, however one has to keep in mind, that for tetrahedral particles, many orientations will be regarded equivalent and some information can be therefore lost.

The $\rho_8$ parameter is defined in the following way:
\begin{equation} \label{eq:octahedral}
    \rho_8(r) = \lim_{dr \to 0} \frac{1}{6}
    \left<
        5 \left[
            \sum_{i,j}\left(
                \vec{u_i} \cdot \vec{v_j}
            \right)^4
        \right] - 9
    \right>_{[r, r+dr]},
\end{equation}
where the summation goes over all $3 \times 3$ pairs $(\vec{u_i}, \vec{v_j})$ of equivalent 4-fold axes of the cube from which the two particles particle were carved. The average is taken for all pairs of particles whose distance is within the $[r, r + dr]$ interval.

\begin{figure}[htb]
    \vspace{0.5in}
    \centering
	\includegraphics[width=0.7\columnwidth]{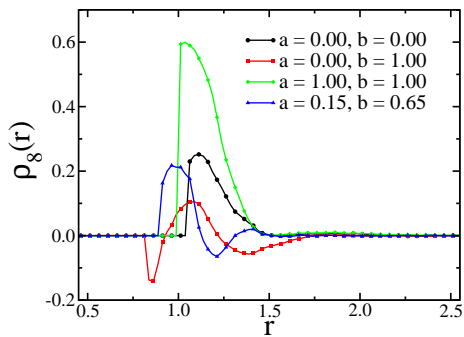}
    \caption{\blue{The dependence of octahedral order parameter $\rho_8(r)$ on the distance between two particles.}}
    \label{fig:order}
\end{figure}
The dependence $\rho_8(r)$ for a few chosen polyhedra is shown in Fig.\ \ref{fig:order}. All parameters decay very quickly with distance, which is typical for RSA \cite{Ciesla2018cubes, Ciesla2018cuboids, Kubala2019}. The highest order is observed for small distances, because the faces of close particles have to be aligned to prevent an overlap, however there is still a certain amount of freedom in rotations around the faces' normal which prevents a full order. If one assumes that all non-overlapping configurations of close particles are equally probable, the numerics show that for polyhedra whose faces closest to the midpoint correspond to the faces of the initial cube, $\rho_8 \approx 0.56$, and for polyhedra whose closest faces are perpendicular to $\vec{a}$ or $\vec{b}$, $\rho_8 \approx 0.35$. While for cubes the value agrees with simulations, for other presented shapes it is significantly larger---tetrahedra even have a negative value. This suggests that some relative orientations of close shapes are blocked by surrounding particles.

\begin{figure}[htb]
    \vspace{0.5in}
    \centering
	\includegraphics[width=0.7\columnwidth]{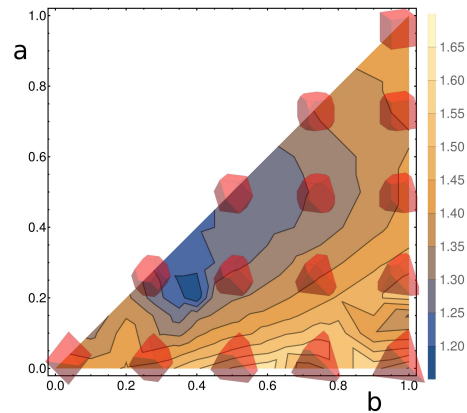}
    \caption{\blue{The dependence of the range of orientational order correlations $\Gamma(a, b)$ on parameters $a$ and $b$.}}
    \label{fig:order_range}
\end{figure}
As an observable for the range of orientational order one can define:
\begin{equation}
    \Gamma = \max\{r : |\rho_8(r)| > \epsilon\},
\end{equation}
where $\epsilon$ is an arbitrary constant. The results for the polyhedra studied are depicted in Fig.~\ref{fig:order_range}. Here, $\epsilon = 0.012$, and the specific value has been chosen so that the function $\Gamma(a, b)$ is as smooth as possible. Octahedral particles have in general a smaller range than proximate tetraheral ones. The minimal value $\Gamma = 1.165$ is reached near $a \approx 0.2$, $b \approx 0.4$ and maximal $\Gamma = 1.655$ for tetrahedron, however due to the way $\Gamma$ was defined, it may be sensitive to statistical fluctuations. The relative difference between the bounds is about $30 \%$, so the range of orientational order does not vary much in $(a, b)$ space. One last remark regarding $\Gamma$ would be that its dependence on $a$, $b$ does not correlate with packing fraction, so there seems to be no straightforward relation between orientational order and packing density.}
\section{Summary}
We have showed that. in general. RSA packings build of solids of octahedral symmetry are less dense than the packings containing solids of tetrahedral symmetry. However, the least dense of the studied packings was build of tetrahedra; its density is $\theta_4 = 0.34750 \pm 0.00049$. A denser value is observed for a packing built of octahedra $\theta_8 = 0.37643 \pm 0.00033$, but the densest packing among  the objects studied here is built of solids similar to the tetrahedron, but with its vertices and edges slightly cut, and equal to $\theta_{max} = 0.41278 \pm 0.00059$. The density autocorrelation function for objects inside the packing is typical for disordered media and vanishes quickly with distance. \blue{The range of orientational order is not significantly influenced by the particle's shape.} An interesting question for future study is to find the random jammed close-packings of these shapes, and see how the behavior found here for RSA correlates with the random jammed packings.
\section*{Acknowledgments}
This work was supported by grant no.\ 2016/23/B/ST3/01145 of the National Science Center, Poland. Numerical simulations were carried out with the support of the Interdisciplinary Center for Mathematical and Computational Modeling (ICM) at the University of Warsaw under grant no.\ G-27-8.
\section*{Appendix A}
Assuming the solids are created by cutting the corners of cube of a unit volume centered at the origin, the coordinates $(x,y,z)$ of the vertices depend on the determinant $\Delta = a + b - 1$ and are given by:
\begin{itemize}
    \item for $\Delta \ge 0$: \\
    All permutations of $\left(\pm (a-\frac{1}{2}), \pm \frac{1}{2}, \pm \frac{1}{2}\right)$ with an even number of minus signs and all permutations of $\left(\pm (b-\frac{1}{2}), \pm \frac{1}{2}, \pm \frac{1}{2}\right)$ with an odd number of minus signs.
    \item for $\Delta \le 0$: \\
    All permutations of $\left(\pm \frac{1}{2}(a+b), \pm \frac{1}{2}(a-b), \pm \frac{1}{2}\right)$ with an even number of minus signs.
\end{itemize}

The virial coefficient $b_2$ equals:%
\begin{equation}
    b_2 = 1 + \frac{R S}{V},
\end{equation}
where $4\pi R$ is the mean curvature integrated over the surface. For convex polyhedra, $R$ can be calculated using the formula \cite{Irrgang2017}:
\begin{equation}
    R = \frac{1}{4\pi}\sum_i l_i\frac{\pi-\theta_i}{2},
\end{equation}
where $l_i$ are the lengths of edges and $\theta_i$ are the corresponding dihedral angles.

Thus, coefficients $c_1 = 2b_2$ can be easily calculated for all studied solids \cite{Boublik1986, Irrgang2017}. Their volumes $V$, surface areas $S$, and curvature $R$ can be expressed as%
\begin{align}
    V &= 1 - \frac{2}{3}\tilde{a}^3 - \frac{2}{3}\tilde{b}^3, \\
    S &= 6 - (6 - 2\sqrt{3})(\tilde{a}^2 + \tilde{b}^2), \\
    R &= \frac{3\sqrt{2}}{4\pi}(1-\Delta) \arccos \left(-\frac{1}{3}\right) + \frac{3}{4}\Delta
\end{align}
for $\Delta \ge 0$ and%
\begin{align}
    V &= 1 - \frac{2}{3}\tilde{a}^3 - \frac{2}{3}\tilde{b}^3 - \frac{1}{2}\Delta^3, \\
    S &= 12ab + 2\sqrt{3}(\tilde{a}^2 + \tilde{b}^2) - 3\sqrt{3}\Delta^2, \\
    R &= \frac{3\sqrt{2}}{4\pi}(1+2\Delta) \arccos \left(-\frac{1}{3}\right) - \frac{3\sqrt{2}}{4}\Delta
\end{align}
for $\Delta \le 0$, where $\tilde{a} = 1 - a$, $\tilde{b} = 1 - b$.

For the tetrahedron, cube, and octahedron, $c_1$ is then given by:
\begin{eqnarray}
    c_1^\mathrm{tet} & = & 2 + \frac{9\sqrt{6}}{\pi}\arccos\left( -\frac{1}{3} \right) \approx 15.407, \nonumber \\
    c_1^\mathrm{cub} & = & 11, \\
    c_1^\mathrm{oct} & = & 2 + \frac{9\sqrt{6}}{\pi}\arccos\left( \frac{1}{3} \right) \approx 10.638. \nonumber 
\end{eqnarray}
The global minimum of $c_1$ is reached for $a=b \approx 0.340$ and is approximately $8.995$.
\section*{Appendix B\label{sec:results_table}}
In Table 
we give our results for the packing fraction $\theta$ for all the values of $a$ and $b$ that we considered, as well as values of the parameters $A$ and $d$ that result from the fit of the time-dependent data to Eq.\ (\ref{fl}).

\begin{table*}[t]
\begin{tabular}{c|c|c|c|c||c|c|c|c|c}
    $a$ & $b$ & $\theta$ & $A$ & $d$ & $a$ & $b$ & $\theta$ & $A$ & $d$ \\
    \hline
    0.00 & 0.00 & 0.37643(33) & 0.2306(84) & 5.87(13) & 0.00 & 0.10 & 0.39113(51) & 0.289(13) & 5.90(15) \\
    0.00 & 0.20 & 0.39501(52) & 0.308(13) & 5.94(15) & 0.00 & 0.30 & 0.39451(51) & 0.320(12) & 6.02(14) \\
    0.00 & 0.40 & 0.39065(44) & 0.3167(92) & 6.14(11) & 0.00 & 0.50 & 0.38546(46) & 0.3015(80) & 6.43(12) \\
    0.00 & 0.60 & 0.37786(45) & 0.2980(77) & 6.47(12) & 0.00 & 0.70 & 0.36940(40) & 0.3098(74) & 6.34(10) \\
    0.00 & 0.80 & 0.36228(61) & 0.2865(91) & 6.66(15) & 0.00 & 0.90 & 0.35399(61) & 0.298(10) & 6.46(16) \\
    0.00 & 1.00 & 0.34750(49) & 0.2935(83) & 6.46(13) & 0.05 & 0.25 & 0.40354(52) & 0.308(14) & 5.79(15) \\
    0.05 & 0.35 & 0.40637(55) & 0.312(13) & 5.95(15) & 0.05 & 0.40 & 0.40763(51) & 0.2908(95) & 6.31(14) \\
    0.05 & 0.45 & 0.40644(53) & 0.315(11) & 6.10(14) & 0.05 & 0.50 & 0.40466(46) & 0.324(10) & 6.01(12) \\
    0.05 & 0.55 & 0.40478(65) & 0.307(12) & 6.33(17) & 0.05 & 0.60 & 0.40250(55) & 0.316(11) & 6.26(14) \\
    0.05 & 0.65 & 0.40017(67) & 0.318(13) & 6.25(17) & 0.05 & 0.70 & 0.39816(66) & 0.307(12) & 6.42(17) \\
    0.05 & 0.75 & 0.39628(65) & 0.295(10) & 6.62(16) & 0.05 & 1.00 & 0.38261(57) & 0.3065(98) & 6.44(14) \\
    0.10 & 0.10 & 0.38594(29) & 0.2219(78) & 5.79(12) & 0.10 & 0.20 & 0.40056(34) & 0.2658(85) & 5.87(11) \\
    0.10 & 0.30 & 0.40719(46) & 0.295(13) & 5.76(14) & 0.10 & 0.35 & 0.40927(47) & 0.299(12) & 5.81(14) \\
    0.10 & 0.40 & 0.41168(47) & 0.2748(94) & 6.20(14) & 0.10 & 0.45 & 0.41157(38) & 0.3027(92) & 5.94(11) \\
    0.10 & 0.50 & 0.41232(56) & 0.288(11) & 6.19(16) & 0.10 & 0.55 & 0.41167(55) & 0.311(13) & 6.01(15) \\
    0.10 & 0.60 & 0.41191(44) & 0.2933(83) & 6.31(12) & 0.10 & 0.65 & 0.41055(53) & 0.297(10) & 6.29(14) \\
    0.10 & 0.70 & 0.40880(52) & 0.311(11) & 6.18(14) & 0.10 & 0.75 & 0.40687(52) & 0.310(10) & 6.21(14) \\
    0.10 & 0.80 & 0.40518(54) & 0.2984(98) & 6.37(14) & 0.10 & 0.90 & 0.40099(60) & 0.2889(96) & 6.57(16) \\
    0.10 & 1.00 & 0.39887(56) & 0.2926(91) & 6.52(14) & 0.15 & 0.25 & 0.40095(36) & 0.262(10) & 5.76(13) \\
    0.15 & 0.35 & 0.40800(43) & 0.2603(99) & 6.00(14) & 0.15 & 0.40 & 0.40935(50) & 0.275(13) & 5.87(16) \\
    0.15 & 0.45 & 0.41154(44) & 0.2724(99) & 6.03(14) & 0.15 & 0.50 & 0.41176(38) & 0.2831(92) & 5.94(12) \\
    0.15 & 0.55 & 0.41210(46) & 0.287(11) & 5.96(14) & 0.15 & 0.60 & 0.41262(44) & 0.2824(93) & 6.13(13) \\
    0.15 & 0.65 & 0.41278(59) & 0.274(11) & 6.32(17) & 0.15 & 0.70 & 0.41171(50) & 0.2789(93) & 6.31(14) \\
    0.15 & 0.75 & 0.40960(46) & 0.2887(92) & 6.19(13) & 0.15 & 1.00 & 0.40287(51) & 0.2902(98) & 6.26(14) \\
    0.20 & 0.20 & 0.38826(34) & 0.2023(95) & 5.74(15) & 0.20 & 0.30 & 0.39892(32) & 0.2578(98) & 5.63(12) \\
    0.20 & 0.35 & 0.40235(34) & 0.2614(99) & 5.68(12) & 0.20 & 0.40 & 0.40585(40) & 0.2446(92) & 6.02(14) \\
    0.20 & 0.45 & 0.40744(42) & 0.256(10) & 5.94(14) & 0.20 & 0.50 & 0.40910(45) & 0.256(10) & 6.04(15) \\
    0.20 & 0.55 & 0.40968(47) & 0.256(10) & 6.09(15) & 0.20 & 0.60 & 0.41003(48) & 0.2547(98) & 6.18(15) \\
    0.20 & 0.65 & 0.40925(37) & 0.2659(83) & 6.05(12) & 0.20 & 0.70 & 0.40954(45) & 0.2539(82) & 6.33(14) \\
    0.20 & 0.75 & 0.40715(36) & 0.2733(80) & 6.06(11) & 0.20 & 0.80 & 0.40593(44) & 0.2655(85) & 6.25(13) \\
    0.20 & 0.90 & 0.40391(46) & 0.2723(94) & 6.16(14) & 0.20 & 1.00 & 0.40160(47) & 0.2720(98) & 6.14(14) \\
    0.25 & 0.25 & 0.38663(31) & 0.209(10) & 5.55(14) & 0.25 & 0.35 & 0.39712(40) & 0.2203(89) & 6.03(15) \\
    0.25 & 0.45 & 0.40250(43) & 0.2294(93) & 6.10(16) & 0.25 & 0.55 & 0.40596(42) & 0.2287(82) & 6.25(15) \\
    0.25 & 0.65 & 0.40632(51) & 0.243(10) & 6.19(17) & 0.25 & 0.75 & 0.40419(53) & 0.245(10) & 6.25(17) \\
    0.25 & 1.00 & 0.39828(47) & 0.274(12) & 5.89(15) & 0.30 & 0.30 & 0.38498(35) & 0.1901(96) & 5.78(17) \\
    0.30 & 0.40 & 0.39332(34) & 0.2225(86) & 5.86(13) & 0.30 & 0.50 & 0.39851(36) & 0.250(11) & 5.68(13) \\
    0.30 & 0.60 & 0.40183(38) & 0.2356(86) & 6.01(14) & 0.30 & 0.70 & 0.40112(45) & 0.2389(96) & 6.11(16) \\
    0.30 & 0.80 & 0.39962(41) & 0.2536(99) & 5.93(14) & 0.30 & 0.90 & 0.39799(48) & 0.239(10) & 6.15(16) \\
    0.30 & 1.00 & 0.39549(35) & 0.2548(83) & 5.96(12) & 0.35 & 1.00 & 0.39257(37) & 0.2441(88) & 5.97(13) \\
    0.40 & 0.40 & 0.38591(41) & 0.204(12) & 5.71(19) & 0.40 & 0.50 & 0.39369(42) & 0.2062(84) & 6.21(16) \\
    0.40 & 0.60 & 0.39625(52) & 0.2092(93) & 6.39(19) & 0.40 & 0.70 & 0.39590(39) & 0.249(10) & 5.80(14) \\
    0.40 & 0.80 & 0.39466(39) & 0.2421(99) & 5.87(14) & 0.40 & 0.90 & 0.39192(34) & 0.2580(97) & 5.72(12) \\
    0.40 & 1.00 & 0.39029(40) & 0.2324(89) & 6.06(14) & 0.45 & 1.00 & 0.38757(44) & 0.242(11) & 5.85(16) \\
    0.50 & 0.50 & 0.38699(36) & 0.214(10) & 5.68(16) & 0.50 & 0.60 & 0.39118(42) & 0.222(11) & 5.89(17) \\
    0.50 & 0.70 & 0.39199(36) & 0.2241(86) & 5.96(14) & 0.50 & 0.80 & 0.39056(40) & 0.2249(92) & 6.00(15) \\
    0.50 & 0.90 & 0.38789(40) & 0.2287(96) & 5.96(15) & 0.50 & 1.00 & 0.38496(40) & 0.247(11) & 5.73(15) \\
    0.55 & 1.00 & 0.38303(41) & 0.2254(96) & 5.98(16) & 0.60 & 0.60 & 0.38534(34) & 0.220(11) & 5.53(15) \\
    0.60 & 0.70 & 0.38633(38) & 0.2102(93) & 5.91(16) & 0.60 & 0.80 & 0.38438(33) & 0.2219(88) & 5.80(13) \\
    0.60 & 0.90 & 0.38131(31) & 0.251(10) & 5.50(12) & 0.60 & 1.00 & 0.37974(30) & 0.2218(73) & 5.91(12) \\
    0.65 & 1.00 & 0.37653(42) & 0.223(11) & 5.78(17) & 0.70 & 0.70 & 0.37861(23) & 0.246(11) & 5.10(11) \\
    0.70 & 0.80 & 0.37742(29) & 0.238(11) & 5.35(13) & 0.70 & 0.90 & 0.37524(26) & 0.2300(85) & 5.53(11) \\
    0.70 & 1.00 & 0.37302(25) & 0.2348(83) & 5.54(11) & 0.75 & 1.00 & 0.37094(35) & 0.2156(97) & 5.77(15) \\
    0.80 & 0.80 & 0.37169(25) & 0.248(12) & 5.08(12) & 0.80 & 0.90 & 0.36962(38) & 0.239(15) & 5.32(17) \\
    0.80 & 1.00 & 0.36760(36) & 0.236(13) & 5.43(15) & 0.85 & 1.00 & 0.36540(40) & 0.228(13) & 5.49(17) \\
    0.90 & 0.90 & 0.36548(32) & 0.241(13) & 5.27(14) & 0.90 & 1.00 & 0.36291(26) & 0.254(11) & 5.21(11) \\
    0.95 & 1.00 & 0.36135(28) & 0.248(12) & 5.26(12) & 1.00 & 1.00 & 0.36030(27) & 0.2356(95) & 5.42(12) \\
    \hline
\end{tabular}
\end{table*}

%
%
%
\bibliography{main}
\end{document}